\documentclass[12pt]{iopart}

\usepackage[T1]{fontenc}
\usepackage{tabularx}
\usepackage{amssymb}
\usepackage{iopams}
\usepackage{bbold}
\usepackage{graphicx}
\usepackage{dcolumn}
\usepackage{bm}
\usepackage{color}

\begin{document}

\title[Tailoring magnetic insulator proximity effects in graphene]{Tailoring magnetic insulator proximity effects in graphene : First-principles calculations}
\date{\today}
\author{A.~Hallal$^{1,2,3}$, F.~Ibrahim$^{1,2,3}$, H.~X.~Yang$^{1,2,3}$, S.~Roche$^{4,5}$ and M.~Chshiev$^{1,2,3}$}
\address{$^1$ Univ. Grenoble Alpes, INAC-SPINTEC, F-38000 Grenoble, France}
\address{$^2$ CEA, INAC-SPINTEC, F-38000 Grenoble, France}
\address{$^3$ CNRS, SPINTEC, F-38000 Grenoble, France} 
\address{$^4$ Catalan Institute of Nanoscience and Nanotechnology (ICN2), CSIC and The Barcelona Institute of Science and Technology, Campus UAB, Bellaterra, 08193 Barcelona, Spain}
\address{$^5$ ICREA-Institució Catalana de Recerca i Estudis Avançats, 08010 Barcelona, Spain}

\begin{abstract}
We report a systematic first-principles investigation of the influence of different magnetic insulators on the magnetic proximity effect induced in graphene. Four different magnetic insulators are considered: two ferromagnetic europium chalcogenides namely EuO and EuS and two ferrimagnetic insulators yttrium iron garnet (YIG) and cobalt ferrite (CFO). The obtained exchange-splitting varies from tens to hundreds of meV. We also find an electron doping induced by YIG and europium chalcogenides substrates, that shift the Fermi level up to 0.78~eV and 1.3~eV respectively, whereas hole doping up to 0.5~eV is generated by CFO. Furthermore, we study the variation of the extracted exchange and tight binding parameters as a function of the EuO and EuS thicknesses. We show that those parameters are robust to thickness variation such that a single monolayer of magnetic insulator can induce a large magnetic proximity effect on graphene. Those findings pave the way towards possible engineering of graphene spin-gating by proximity effect especially in view of recent experiments advancement.
\end{abstract}
\maketitle

\section {Introduction}
 
Graphene spintronics is one of the most promising directions of innovation for two-dimensional materials, opening new prospects for information technologies~\cite{Roche, Han1}. 
Besides its exceptional electrical, thermal, and mechanical properties~\cite{Neto, Geim}, two-dimensional graphene possesses a unique electronic band structure of massless Dirac fermions with a very long spin-diffusion lengths up to room temperature owing to its weak intrinsic spin-orbit coupling~\cite{Tombros, Popinciuc, Dlubak, Han, YYang, Massen, Dlubak2,Cummings,VanTuan}. Accordingly graphene stands as a potential spin-channel material. However, a fundamental challenge lies in the development of external ways to control the propagation of spin currents at room temperature, in view of designing spin logics devices.

Since carbon is non-magnetic, a significant effort is focused on injecting spins and inducing magnetism in graphene. 
Magnetism in graphene can be induced and controlled through material design or defects and several methods have been proposed to magnetize graphene~\cite{Yazyev1,Yazyev2}.
For instance,  edge magnetism has been shown to develop in a few nanometers wide graphene nanoribbons for certain edge geometries~\cite{Son, Kim}, or the hole structure of graphene nanomesh~\cite{Bai} was also theoretically proposed to offer robust and room temperature magnetic states able to affect spin transport~\cite{Yang2,Pedersen}. Much attention is also paid to tailor spin-polarized currents and magnetoresistance signals by intentional defects, or depositing atoms~\cite{Sariano, McCreary, Chan, Qiao, Ding, Zhang, Jiang} or molecules~\cite{Park, Kim2, JWYang}. Recently, the production of spin-polarized currents and magnetoresistance signals by growing graphene on magnetic substrates, such as YIG, has raised a lot of interest~\cite{Wang, Leutenantsmeyer, Evelt,Singh}. However, the conductivity mismatch is an important factor that influences the spin injection from magnetic metallic substrates into graphene restricting, thus, the design of novel types of spin switches.  Therefore, the use of magnetic insulators (MIs) has attracted much interest as an alternative route to induce magnetism in graphene via the exchange-proximity interaction.
 
Prior theoretical study of proximity effects of a ferromagnetic insulator (EuO) on graphene reported a large spin polarization of $p$ orbitals together with a large exchange-splitting band gap~\cite{Yang}. However, the drawback of using EuO is its low Currie Temperature (T$_C$) and the predicted strong electron doping (about 1.4~eV). Thus, many theoretical works have been dedicated to investigate different MIs with higher T$_C$ and weaker doping which is crucial for practical spintronic devices. Additionally, theoretical investigations of graphene in proximity of topological~\cite{Qiao2,Vobornik} and  multiferroic insulator~\cite{Zanolli} have found large exchange-splitting up to 300~meV. Recently, it has been proposed to insert 2D insulators (e.g. hBN) between graphene and the ferromagnetic material to induce exchange splitting~\cite{Zollner}. In this case, the doping of graphene and exchange coupling strengh can be tuned by varying the number of hBN layers. On the other hand, recent experiments on YIG/Gr~\cite{Wang, Leutenantsmeyer, Evelt,Mendes} and EuS/Gr~\cite{Wei} demonstrated a large exchange-coupling between MI and graphene. Namely, a large magnetic exchange field up to 14~T is found in case of EuS on graphene with a potential of reaching hundreds of Tesla. However, EuS has even lower T$_C$ compared to that of EuO. For YIG/Gr some experiments show a very large exchange-coupling of the order of tens of meV~\cite{Wang} while others reported smaller values of 0.2~T~\cite{Leutenantsmeyer, Evelt} or 1~T~\cite{Singh}. Such discrepancy between the two reports might be due to different absorption strengths between graphene layer and YIG.

In this Letter, using first-principles calculations we explore how the nature of the magnetic insulator affects the features of the magnetic proximity effect induced in graphene. Four cases of different MIs are studied: europium oxide (EuO), europium sulfite (EuS), cobalt ferrite CoFe$_{2}$O$_{4}$ (CFO) as well as yttrium iron garnet Y$_{3}$Fe$_{5}$O$_{12}$ (YIG). The exchange-proximity parameters are obtained from the electronic-band structure of graphene calculated in each case. We obtain electron doping for all cases except the CFO where the Dirac point lies above the Fermi level at 0.5~eV. The magnetic proximity effect results in a large exchange-splitting band gap of a few tens of meV. The presence of spin-dependent band gap around Dirac point is clear in all cases, except for cobalt ferrite where no bandgap is formed. In addition, we report systematic studies of electronic band structure of graphene as a function of EuO and EuS thickness where we show that the exchange-splitting gaps are robust to MI thickness variation. These findings pave the way towards possible engineering of graphene spin-gating by proximity effect especially in view of aforementioned recent experiments on EuS and YIG on top of graphene.

\begin{table*} [t]
\caption{Computational and Structural details for the four investigated systems, effective Hubbard term, the bulk lattice parameter, the lattice mismatch between the MIs and graphene and the Curie temperature of each magnetic insulators.}
\footnotesize
\begin{tabular}{@{}lcccccr}
   \br
   Structure&Package&Potential&$U_{eff}$(eV)&Latt. param.(\AA)&Mismatch(\%)&Tc(K)\\
   \mr
   EuO & SIESTA & LDA+U & Eu-$f$ 7.6 O-$p$ 3.4 & 5.18 & 0.8 &  77 \\
   EuS & SIESTA & LDA+U & Eu-$f$ 6.3 and Eu-$d$ 4.4 &  5.92 &  -1.76 & 16.5\\
   Y$_3$Fe$_5$O$_12$ & SIESTA & LDA+U &  Fe-$d$ 2.7  & 12.49 & 2.5 & 550  \\
   CoFe$_2$O$_4$ & VASP & GGA+U &  Fe-$d$ 3.61 Co-$d$ 3.61 & 8.46 &  -3.6 & 793\\
\br
\end{tabular}

  \label{tab:1}
\end{table*}

\section{Methodolgy}
The Vienna ab initio simulation package (VASP) \cite{Kresse1,Kresse2,Kresse3} is used for structure optimization, where the electron-core interactions are described by the projector augmented wave method for the potentials \cite{Blochl}, and the exchange correlation energy is calculated within the generalized gradient approximation (GGA) of the Perdew-Burke-Ernzerhof form \cite{Perdew, Kresse4}. The cutoff energies for the plane wave basis set used to expand the Kohn-Sham orbitals are 500~eV for all calculations. Structural relaxations and total energy calculations are performed ensuring that the Hellmann-Feynman forces acting on ions are less than $10^{-2}$ for all studied structure. Except for YIG, due to its large supercell, relaxation is done using SIESTA code \cite{Soler}, where the exchange correlation energy is calculated within the local density approximation (LDA)~\cite{Ceperley,Perdew2}. 

Since Eu is a heavy element with atomic number $63$ and its outer shell ($4f^7$ $6s^2$) contains $4f$ electrons, the GGA and LDA approaches fail to describe the strongly correlated localized $4f$ electrons and predicts a metallic ground state for the europium chalcogenides, whereas a clear band gap is observed in experiments. Similarly, GGA and LDA fail to describe the electronic interaction in Mott insulator such as iron oxides or cobalt oxides. Such a deficiency of these approaches is expected in correlated systems as transition metal oxides. Thereby, to account for the strong on-site Coulomb repulsion among the localized $3d$ ($4f$) electrons in YIG and CFO (EuO and EuS) we have introduced a Hubbarad-U parameter as described by the authors of Ref.~\cite{Anisimov1, Anisimov2} for SIESTA and as described and implemented in the VASP code. The LDA+U and GGA+U represented by the Hubbard-like term $U$ and the exchange term $J$, which led to an improvement of the ground state properties such as the energy band gap and the spin magnetic moments in the MIs. The $U_{eff}=U-J$ value used for each system is summarized in \Tref{tab:1}, and in addition to $U_{eff}$ for Eu-$f$ in EuO, the LDA in EuS is also corrected by adding $U_{eff}$ term to the Eu-$d$ orbitals following Ref.~\cite{Larson}.
In all cases investigated, the density of states of bulk MIs are calculated and compared to those obtained using the VASP package and a good agreement is found between the two approaches using the same U parameters.

\begin{figure*}[t]
  \centering
     \includegraphics[width=0.9\textwidth]{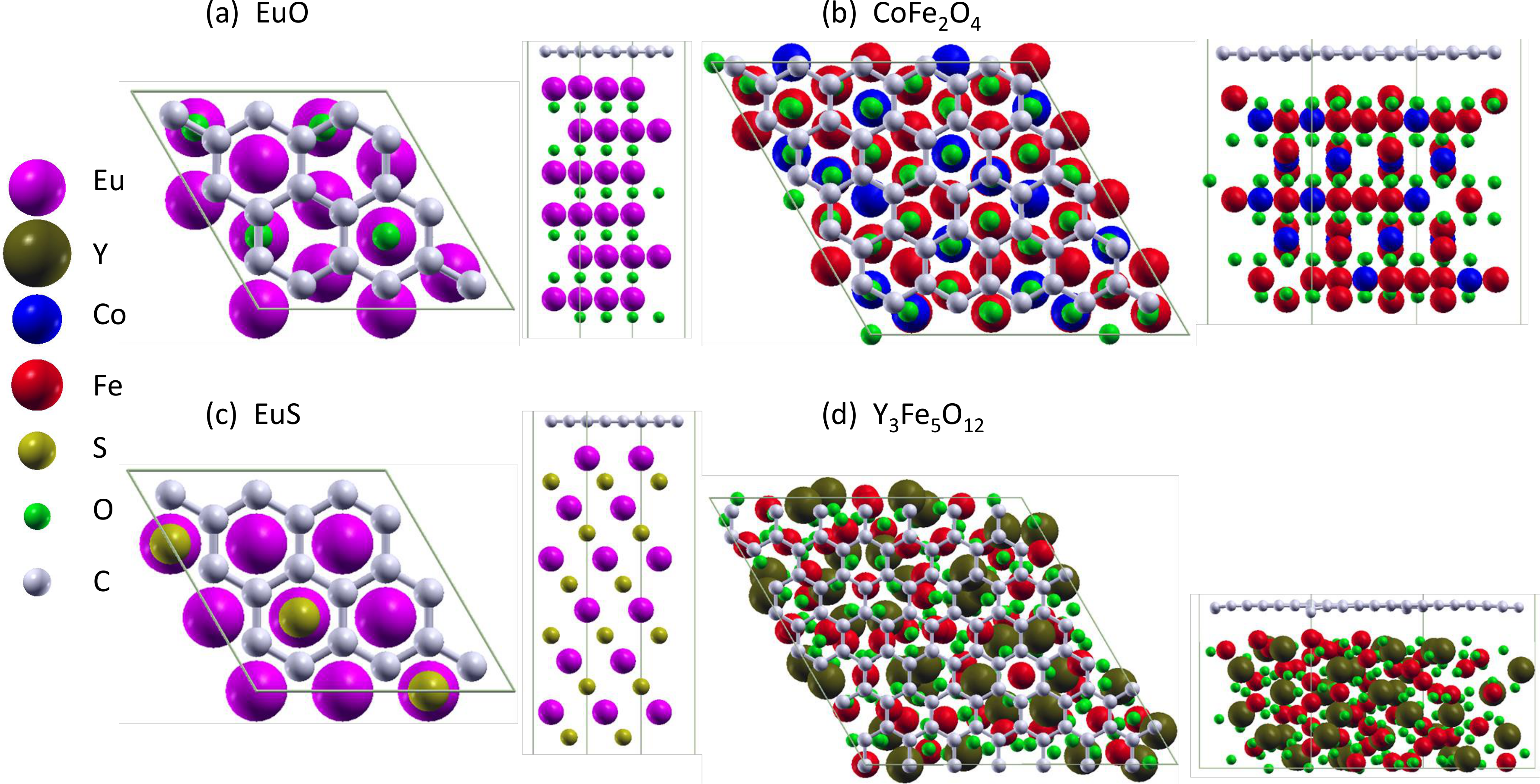}
  \caption{(Color online)  Side view and top view of the
calculated crystalline structures for graphene on top of
(a) EuO film (b) CoFe$_{2}$O$_{4}$(C) EuS (d) Y$_{3}$Fe$_{5}$O$_{12}$. 
All the calculated structure are passivated with with hydrogen atoms.}
  \label{fig1}
\end{figure*} 

The two investigated EuO and EuS compounds have a ferromagnetic ground state with a rocksalt structure with lattice parameters of 5.18~\AA{} and 5.92~\AA{}, respectively. Lattice structure and lattice mismatch between graphene and EuO are described in detail in Ref.~\cite{Yang}. 
It is found that a 3$\times$3 unit cell of graphene can fit easily on a a 2$\times$ 2 EuO (111) surface unit cell with a lattice about 7.33~\AA{} and with a lattice mismatch of about 0.8\%. For EuS, the bulk lattice parameter is quite larger than that of bulk EuO. Nevertheless, graphene can still fit on a EuS $\sqrt{3}\times\sqrt{3}$ (111) substrate with a lattice mismatch in order of 1.8\%. Due to this difference in the lattice parameter between EuO and EuS, a different graphene absorption on top of the surface occurs as seen in \Fref{fig1} [(a) and (c)]. In both cases the supercell is composed of six bilayer of europium chalcogenides with graphene in top of Eu termination, which is the energetically most stable configuration.

Next, we consider the lattice mismatch between graphene and YIG. Their lattice parameters are 2.46~\AA{} and about 12.49~\AA{}, respectively. as shown in \Fref{fig1}(d), the 1 $\times$ 1 unit cell of YIG (111) substrate with a lattice constant of about 17.66~\AA{} can fit on the 7$\times$7 graphene unit cell, with a lattice mismatch of about 2.5\%.  The resulting supercell is composed of six YIG trilayers and a graphene layer placed on top.
For CFO the bulk lattice parameter is 8.46~\AA{} and again along the (111) direction a 5$\times$5 graphene unit cell can fit on 1$\times$1 CFO(111) substrate with a lattice mismatch of about 3.6\%. The supercell in this case is composed of six trilayer of CFO with graphene in top of Fe atoms (cf.~\Fref{fig1}(b)). In all the cases, the bottom surface is passivated with hydrogen atoms in order to avoid the bottom surface effect on graphene and the vacuum region is chosen to be larger than 14~\AA{}. 
The lattice structure of graphene/MIs are presented in figure~\Fref{fig1} with a vertical distance between Eu and C layers around 2.57~\AA{} and 2.52~\AA{} for EuO and EuS, respectively. For graphene/YIG and graphene/CFO, due to the large lattice mismatch, the graphene lattice is corrugated with corrugation height in order of 0.6~\AA{} and 0.15~\AA{} for YIG and CFO, respectively. The average vertical distance between Fe and C atoms is close to 2.7~\AA{} for both YIG/graphene and CFO/graphene. This strong corrugation presents in graphene may affect its electronic band structure as shown previously for graphene on top of MgO substrate~\cite{Cho}.

Finally, using the SIESTA package and the optimized structures of graphene on MIs shown in \Fref{fig1}, we calculate the electronic structure of the systems with LDA+U for the exchange correlation functional (c.f. \Tref{tab:1}). The self-consistent calculations are performed with an energy cutoff of 600 Ry and with a 4$\times $4 $\times $1 K-point grid for EuO and EuS and 3$\times $3 $\times $1 for YIG. A linear combination of numerical atomic orbitals with a double-$\zeta$  polarized basis set is used for the small structures and and a single-$\zeta$ for the larger ones. For graphene on CFO, the the electronic structure is calculated using  GGA+U as implemented in VASP package with a 3$\times $3 $\times $1 K-point grid.
 
\begin{figure*}[t]
  \centering
     \includegraphics[width=0.9\textwidth]{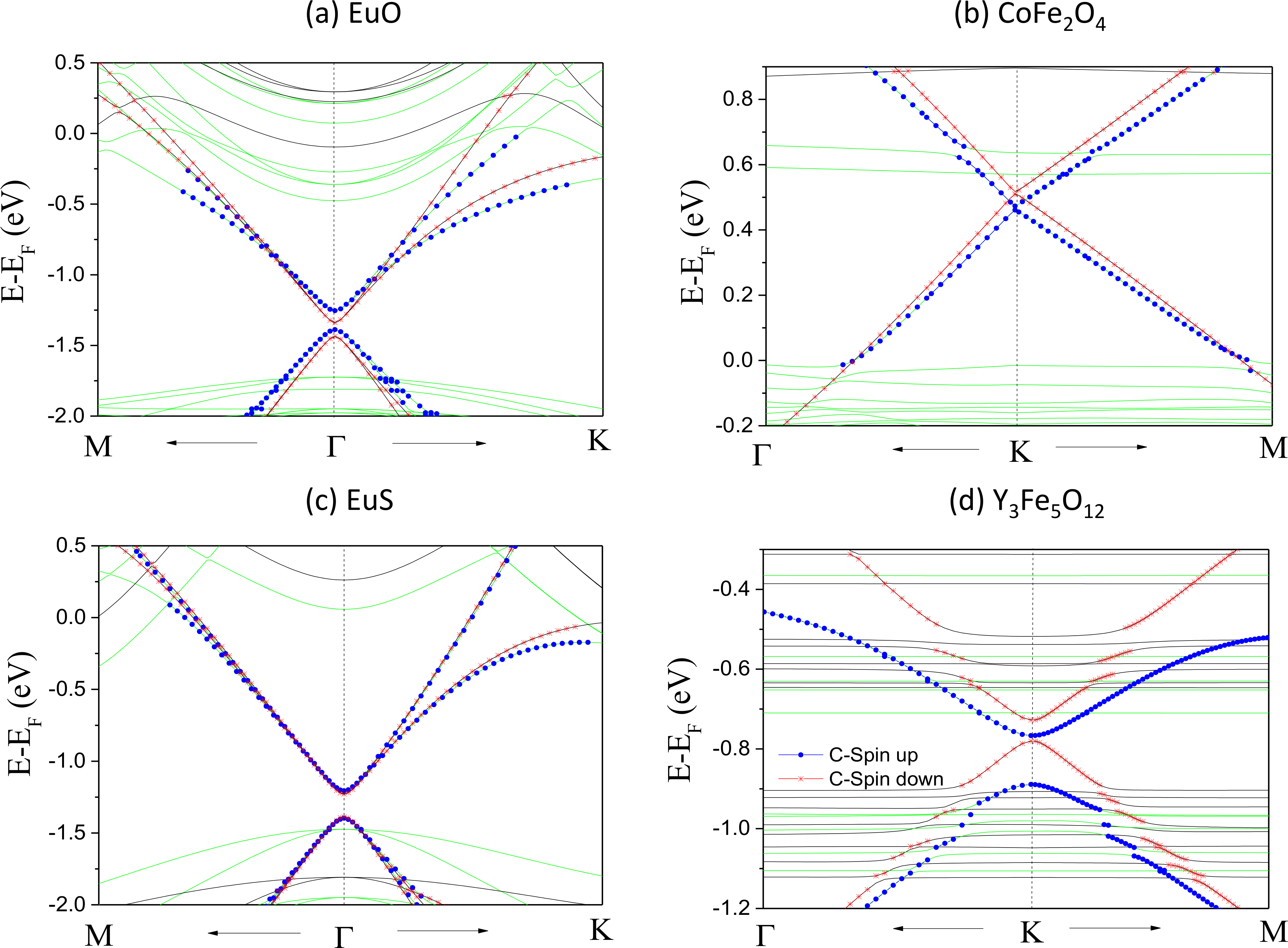}
  \caption{(Color online) Band structures of graphene on (a) EuO (b) CoFe$_{2}$O$_{4}$ (c) EuS (d) Y$_{3}$Fe$_{5}$O$_{12}$.
Blue (Green) and red (black) represent spin up and spin down
bands of graphene (magnetic insulator), respectively. EuO case is taken from Ref.~\cite{Yang}}
  \label{fig2}
\end{figure*}

\section {Results and Discussion}
\subsection{Electronic structure of graphene in proximity of MIs}
Graphene honeycomb structure comprising two equivalent carbon sublattices $A$ and $B$ is responsible for the fact that  charge carriers are described by massless Dirac excitations. Of particular importance for the physics of graphene are the two Dirac points $K$ and $K'$ at the inequivalent corners of the graphene Brillouin zone (BZ). In the vicinity of these two points, the electronic structure of graphene is characterized by a linear dispersion relation with a Dirac point separating the valence and conduction bands with a zero band gap as follows:

\begin{equation}
 H_{1}(q)=  v_F \bsigma \bdot \bi{q}
\end{equation}

where $q$ is the momentum measured relative to the Dirac point and $v_F$ represents the Fermi velocity which does not depend on the energy or momentum~\cite{Neto}. The gapless Dirac cones at $K$ and $K'$ are protected by time-reversal and inversion symmetry. Since Dirac points are separated in the BZ, small perturbations cannot lift this valley degeneracy.  Once graphene is in proximity of a substrate, $A$ and $B$ sublattices feel different chemical environment which leads to the inversion symmetry breaking between $K$ and $K'$ and giving rise to a band gap. This can be modeled by the following Hamiltonian describing the graphene's linear dispersion relation in proximity of magnetic insulator:
\begin{equation}\label{Eq2}
 H_{2}(q)=  v_F \bsigma \bdot \bi{q} \mathbb{1}_s + \delta  \mathbb{1}_\sigma  s_z + \frac{\Delta_s}{2} \sigma_z \mathbb{1}_s 
\end{equation}
where $\sigma$ and $s$ are the Pauli matrices that act on sublattice and spin, respectively. The second term represents the exchange coupling induced by the magnetic moment of magnetic atoms, with $\delta$ being the strength of exchange spin-splitting of the hole or electron. The last term results from the fact that graphene sublattices $A$ and $B$ are now feeling different potential which might result in a spin-dependent band gap opening at the Dirac point and $\Delta_s$ is the spin-dependent staggered sublattice potential. A  Rashba spin orbit coupling term can also be added to the Hamiltonian and can be represented by $\frac{\alpha}{2} (\bsigma \wedge \bi{s})$ at the left side of~\Eref{Eq2}.

Let us now discuss the electronic band structures of graphene in proximity to MIs as shown in~\Fref{fig2}. For graphene on top of europium chalcogenides a $3\times 3$ unit cell is used and due to zone folding of graphenes BZ, both $K$ and $K'$ valleys get mapped to the $\Gamma$ point~\cite{Yang}. Therefore for EuO and EuS, the Dirac cone of graphene becomes located at the $\Gamma$ point instead of $K$ one's. The linear dispersion of the graphene band structure is modified with a band gap opening at the Dirac point. More interestingly, this degeneracy lifting at the Dirac point is spin dependent as demonstrated for EuO~\cite{Yang}. The spin-dependent band gaps found in the EuO/graphene are about 134 and 98 meV for spin up and spin down states, respectively~(see \Fref{fig2}(a)). Here, however, we fit the band structure parameters according to Hamiltonian given by \Eref{Eq2} to which the exchange splitting gaps of 84 and 48 meV are added for electrons and holes, respectively. Replacing EuO by EuS increases drastically the band gap opening as shown in \Fref{fig2}(c). The spin-dependent band gaps in this case are about 192 (resp. 160 meV) for spin up (resp. spin down) states. However, the spin splitting is strongly reduced to 23 (resp. -10 meV) for electrons (resp. holes). This difference between EuO and EuS results from the fact that all 3 Eu atoms in EuS case are sitting in a hollow site of graphene hexagon while for EuO, the atoms belong to the bridge site and to the hollow site as shown in \Fref{fig1}(a) and (c). Recently, Su et al~\cite{Su} reported that while Eu atom sitting at the hollow site of graphene hexagon is described by an inter-valley scattering term in the induced proximity Hamiltonian, Eu atoms at the  bridge site reduces the graphene lattice symmetry and can be represented by a valley pseudospin Zeeman term in $x$-direction in sublattice space that shifts slightly the Dirac cones from the $\Gamma$ point.

\begin{table*}[t]
 \caption{Extracted energy gaps and exchange parameters of graphene/MIs structures at Dirac point compared with parameters for graphene in proximity of EuO heterostructure reported in Ref.~\cite{Su}. E$_G$ is the band-gap of the Dirac cone.  $\Delta_\uparrow $ and $\Delta_\downarrow $ are the spin-up and spin-down gap, respectively. The spin-splitting of the electron and hole bands at the Dirac cone are  $\delta_e$ and $\delta_h$ , respectively. E$_D$ is the Dirac cone doping with respect to Fermi level.}
 \footnotesize
\begin{tabular}{@{}lcccccc}
   \br
   Structure & E${_G}$ (meV) & $\Delta_\uparrow $  (meV) & $\Delta_\downarrow  $ (meV) & $\delta_e$ (meV) & $\delta_h$ (meV) & E$_D$ (eV) \\
   \mr
    EuO/Gr/EuO(1BL) aligned\cite{Su} & 127 & 309 & 344 & 182 & 217 & -2.8\\
    EuO/Gr/EuO(1BL) misaligned\cite{Su} & -38 & 137 & 182 & 211 & 220 &-2.8 \\
	GR/EuO(6BL) & 50 & 134 & 98  & 84 & 48 &  -1.37 \\
    GR/EuS(6BL)  & 160 & 192 & 160 & 23 &  -10 & -1.3\\
	Gr/Y$_3$Fe$_5$O$_{12}$ & 1 & 116 &  52  & -52 & -115 & -0.78  \\
	Gr/CoFe$_2$O$_4$ & -37 & 12 &  8 & -45 &  -49 & +0.49 \\
   \br
\end{tabular}

  \label{tab:2}
\end{table*}

Let us now discuss the proximity effects induced by yttrium garnet (YIG) and cobalt ferrite (CFO) oxides. In ~\Fref{fig2}(d) we present the electronic bands of the YIG/Graphene structure where we see that the proximity of YIG induces a band gap opening around the Dirac point. Furthermore, due to the interaction between graphene and the magnetic substrate, the spin-degeneracy around Dirac point is lifted. The spin-dependent band gaps found in the YIG/graphene are 116 and 52 meV for spin up and spin down states, respectively. The spin splittings estimated from the band structure are found to be about -52 and -115 meV  for electrons and holes, respectively. Due to its strong interaction with the magnetic insulator, graphene becomes heavily doped and the Dirac Cone is shifted  below the Fermi level as seen in Figure1 (b). Interestingly, the band structure presented in \Fref{fig1}(b) shows that graphene on top of YIG has a half metallic behavior. The spin-up Dirac cone lies in the middle of the spin-down gap and vice versa.  For the CFO/graphene case the induced  band gap opening around the Dirac point is absent (see Figure~\ref{fig2}(b)). This is due to the quite large interlayer distance between the ferrimagnetic insulator and graphene and due to the physisorption interaction which  does not perturb the inversion symmetry of the two Dirac points. Nevertheless, due to the interaction between graphene and the magnetic substrate, the spin-degeneracy around Dirac point is lifted and spin-dependent band gaps are still induced in this case and found about 12 and 8 meV for spin up and spin down states, respectively. The strength of the exchange-splitting estimated from the band structure is -45 and -49 meV for electrons and holes, respectively. Due to the weak interaction with the magnetic insulator graphene becomes slightly doped and the Dirac Cone is shifted  above the Fermi level as seen in \Fref{fig2}(b). 

The extracted energy band gaps and exchange-splitting values at Dirac point induced in graphene by the proximity of magnetic insulators are summarized in ~\Tref{tab:2} with E$_G$, $\Delta_{\uparrow}$ and $\Delta_{\downarrow}$  representing the energy band gap and the spin-dependent gaps for spin-up and spin-down, respectively. The spin splitting of the electron and hole bands are denoted as $\delta_e$ and $\delta_h$. Finally, E$_D$ indicates how large is the Dirac point doping with respect to Fermi energy. In Table~\ref{tab:2} the positive value of E$_G$ indicates a band gap between conduction and valence band, whereas negative value indicates a spin resolved band overlap as seen in CFO case shown in \Fref{fig2}(b). Spin-splittings are defined by spin dependent energy differences at Dirac point with negative value indicating that spin-up bands are lower in energy than that of spin-down bands. The extracted values are compared with that aligned and misaligned EuO heterostructure with graphene between two EuO layers reported in Ref.~\cite{Su}. As illustrated in Table~\ref{tab:2}, doping graphene with EuO will push further the Dirac point below the Fermi level and makes impossible to harvest the graphene linear dispersion in practical electronic devices.  To overcome the problem of strong doping one can deposit on the top side of the structure a material which can hole dope graphene. For instance, we propose that CFO deposited on the top side of europium chalcogenides/graphene or even YIG/graphene will bring Dirac cone closer to Fermi level and the exchange-splitting parameter induced by proximity effect, in such a heterostructure, is expected to double to be in the range of hundreds of meV. Moreover, this type of asymmetric heterostructure will break the in-plane inversion symmetry of the graphene layer and might give rise to topological properties such as quantum anomalous Hall effect~\cite{Su}. 

\begin{figure*}[t]
  \centering
     \includegraphics[width=0.9\textwidth]{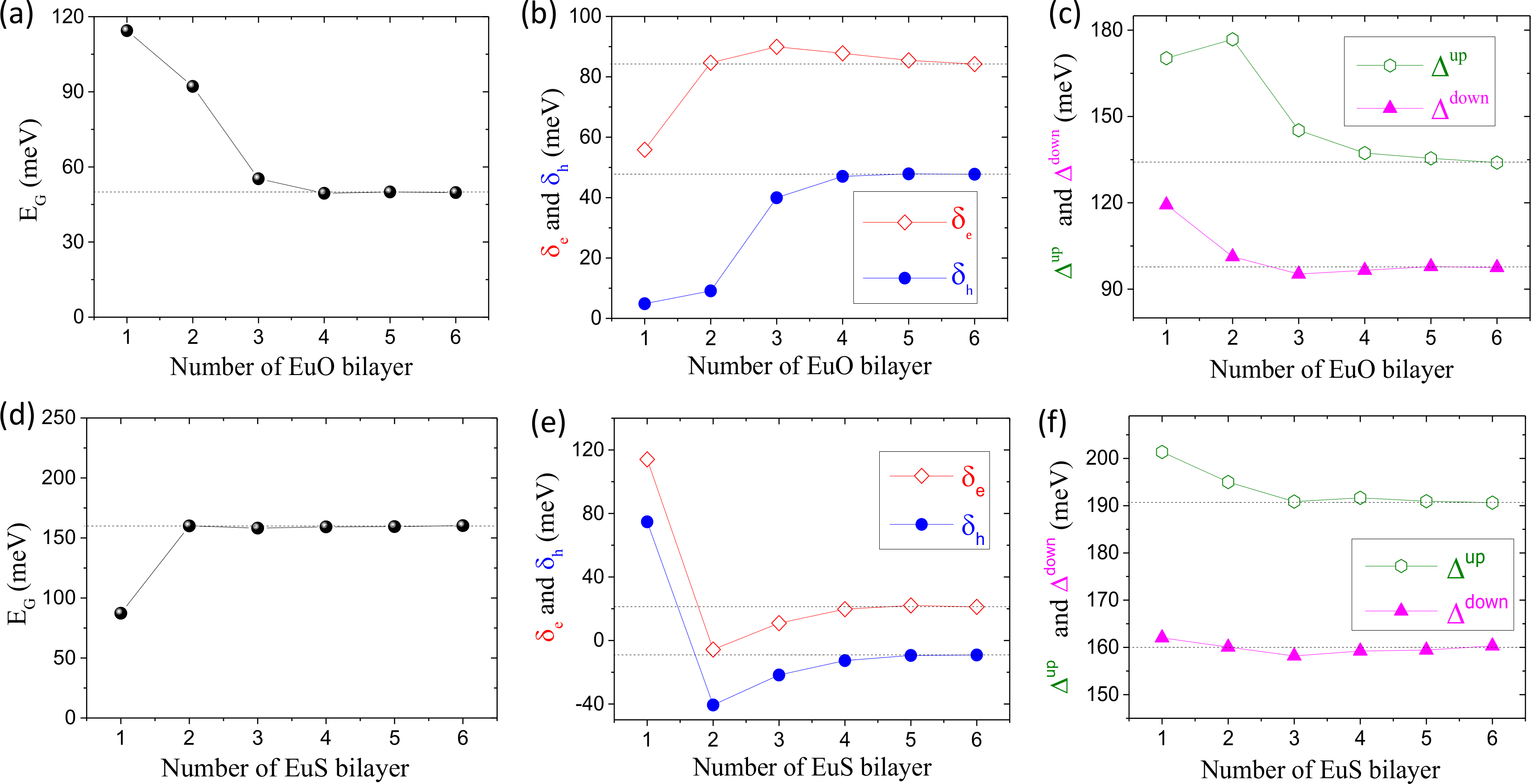}
  \caption{Thickness variation of the exchange-coupling parameters presented in table~\ref{tab:2} for the graphene in proximity of chalcogenides EuO and EuS.}
  \label{fig3}
\end{figure*}

\subsection{Thickness variation effect on the graphene exchange parameter in proximity of europium chalcogenides}

Finally, let us check the robustness of aforementioned results by exploring the variation of the energy band gaps and proximity exchange-splitting in graphene at Dirac point as a function of MIs thickness. As seen in \Fref{fig3}, all the plotted values tends to saturate above a thickness of 3 bilayers indicating that already 3 or 4 bilayers of MIs are sufficient to mimic the bulk effect. The results also indicate that already MIs as thin as 1 bilayer of europium chalcogenides can induce large proximity effect in graphene. For instance, the spin-splitting of the electron and hole bands at the Dirac cone in the case of one bilayer of EuS (EuO) are found about 120 and 80 meV (55 and 5 meV), respectively. As EuS (EuO) thickness increases, both spin-splitting values decreases (increases) to reach the bulk value shown in ~\Tref{tab:2}. As for spin-dependent band gaps $\Delta_{\uparrow}$ and $\Delta_{\downarrow}$, both decreases as a function of MI thickness with variation of spin-down and spin-up band gaps being less dramatic in the case of EuS compared to that for EuO. Since the induced magnetism in graphene due to proximity of europium chalcogenides arises from graphene hybridization with polarized Eu-$4f$ state right below the Fermi level~\cite{Yang}, the observed variation at low thicknesses is related to the variation of the energy level of these Eu-$4f$ states.

\section{Conclusion}

In summary, using first-principles calculations we investigated proximity effects induced in graphene by magnetic insulators. Four different MIs have been considered: two ferromagnetic europium chalcogenides and two ferrimagnetic  insulators yttrium iron garnet and  cobalt ferrite. In all cases, we find that the exchange-splitting varies in the range of tens to hundreds meV. While Dirac cone is negatively doped for europium chalcogenides and YIG, it is found to be positively doped for CFO substrate. In order to bring the Dirac cone closer to the charge neutrality point, we propose to deposit on the top side of the negatively doped structure a material which can positively dope graphene, such as CFO. In such a heterostructure the exchange-coupling parameter induced by proximity effect is expected to be doubled. Moreover, we explored the variation of the extracted magnetic exchange parameters as a function of europium chalcogenides thicknesses. This analysis show that the extracted parameters are robust to thickness variation and one monolayer of magnetic insulator can induce a large magnetic proximity effect on graphene. These findings pave the way towards possible engineering of graphene spin-gating by proximity effect especially in view of recent experiments advancement. 

\ack

This project has received funding from the European Union's Horizon 2020 research and innovation programme under grant agreement No. 696656 (Graphene Flagship). S. R. acknowledges Funding from the
Spanish Ministry of Economy and Competitiveness and the European Regional Development Fund (Project
No. FIS2015-67767-P (MINECO/FEDER)), the Secretaria de Universidades e Investigación del Departamento de Economía y Conocimiento de la Generalidad de Cataluña, and the Severo Ochoa Program (MINECO, Grant No. SEV-2013-0295).


\section*{References}

\begin{thebibliography}{200}
\bibitem{Roche} S. Roche, J. \AA{}kerman, B. Beschoten, J.-C. Charlier, M. Chshiev, S. P. Dash, B. Dlubak, J. Fabian, A. Fert, M. Guimar\~aes, F. Guinea, I. Grigorieva, C. Sch\"onenberger, P. Seneor, C. Stampfer, S. O. Valenzuela, X. Waintal and B. van Wees,  2D Materials  {\bf 2}, 030202 (2015)

\bibitem{Han1} W. Han, R. K. Kawakami, M. Gmitra and J. Fabian, Nature Nanotechnology {\bf 9}, 794 (2014).


\bibitem{Neto} A. H. Castro Neto, F. Guinea, N. M. R. Peres, K. S. Novoselov, and A. K. Geim, Rev. Mod. Phys. {\bf 81}, 109 (2009).
\bibitem{Geim} A. K. Geim and K. S. Novoselov, Nat. Mater. {\bf 6}, 183 (2007).

\bibitem{Tombros} N. Tombros, C. Jozsa, M. Popinciuc, H. T. Jonkman, and B. J. van Wees, Nature (London) {\bf 448}, 571 (2007).

\bibitem{Popinciuc} M. Popinciuc, C. J\'ozsa, P. J. Zomer, N. Tombros, A. Veligura, H. T. Jonkman, and B. J. van Wees, Phys. Rev. B {\bf 80}, 214427 (2009).

\bibitem{Dlubak} B. Dlubak, P. Seneor, A. Anane, C. Barraud, C. Deranlot, D. Deneuve, B. Servet, R. Mattana, F. Petroff, and A. Fert, Appl. Phys. Lett. {\bf 97}, 092502 (2010).

\bibitem{Han} W. Han and R. K. Kawakami, Phys. Rev. Lett. {\bf 107}, 047207 (2011).

\bibitem{YYang} T. -Y. Yang, J. Balakrishnan, F. Volmer, A. Avsar, M. Jaiswal, J. Samm, S. R. Ali, A. Pachoud, M. Zeng, M. Popinciuc, G. G\"untherodt, B. Beschoten, and B. \"Ozyilmaz, Phys. Rev. Lett. {\bf 107}, 047206 (2011)

\bibitem{Massen}  T. Maassen, J. J. van den Berg, N. IJbema, F. Fromm, T. Seyller, R. Yakimova, and B. J. van Wees, Nano Lett. {\bf 12}, 1498 (2012).

\bibitem{Dlubak2} B. Dlubak, M. -B. Martin, C. Deranlot, B. Servet, and S. Xavier, Nat. Phys. 8, 557 (2012).

\bibitem{Cummings} Cummings AW, Roche S , Physical Review Letters 116, 086602 (2016).

\bibitem{VanTuan} Van Tuan D., Ortmann F., Cummings A.W., Soriano D., Roche S. Scientific Reports; 6 (21046) 2016.

\bibitem{Yazyev1} O. V. Yazyev and L. Helm, Phys. Rev. B 75, 125408 (2007); O. V. Yazyev, Phys. Rev. Lett. 101, 037203 (2008).

\bibitem{Yazyev2} O. V. Yazyev, Rep. Prog. Phys. 73, 056501 (2010)

\bibitem{Son} Y.-W. Son, M. L. Cohen, and S. G. Louie, Nature {\bf 444}, 347 (2006).

\bibitem {Kim} W. Y. Kim and K. S. Kim, Nat. Nanotechnol. {\bf 3}, 408 (2008).

\bibitem{Bai} J. Bai, X. Zhong, S. Jiang, Y. Huang, and X. Duan, Nat. Nanotechnol. {\bf 5}, 190 (2010).

\bibitem{Yang2} H. X. Yang, M. Chshiev, D. W. Boukhvalov, X. Waintal, and S. Roche, Phys. Rev. B {\bf 84}, 214404 (2011).

\bibitem{Pedersen} M. L. Trolle, U. S. Møller, and T. G. Pedersen, Phys. Rev. B 88, 195418 (2013).

\bibitem{Sariano} D. Soriano, N. Leconte, P. Ordejon, J.-Ch. Charlier, J.-J. Palacios, and S. Roche, Phys. Rev. Lett. {\bf 107}, 016602 (2011).

\bibitem{McCreary} K. M. McCreary, A. G. Swartz, W. Han, J. Fabian, and R. K. Kawakami, Phys. Rev. Lett. {\bf 109}, 186604 (2012).

\bibitem{Chan} K. T. Chan, J. B. Neaton, and M. L. Cohen, Phys. Rev. B {\bf 77}, 235430 (2008).

\bibitem{Qiao} Z. Qiao, S. A. Yang, W. Feng, W.-K. Tse, J. Ding, Y. Yao, J. Wang, and Q. Niu, Phys. Rev. B {\bf 82}, 161414(R) (2010).

\bibitem{Ding} J. Ding, Z. Qiao, W. Feng, Y. Yao, and Q. Niu, Phys. Rev. B {\bf 84}, 195444 (2011).

\bibitem{Zhang} H. Zhang, C. Lazo, S. Blugel, S. Heinze, and Y. Mokrousov, Phys. Rev. Lett. {\bf 108}, 056802 (2012).

\bibitem{Jiang} H. Jiang, Z. Qiao, H. Liu, J. Shi, and Q. Niu, Phys. Rev. Lett. {\bf 109}, 116803 (2012).

\bibitem{Park} J. Park, S. B. Jo, Y.-J. Yu, Y. Kim, J. W. Yang, W. H. Lee, H. H. Kim, B. H. Hong, P. Kim, K. Cho, and K. S. Kim, Adv. Mater. {\bf 24}, 407 (2012).

\bibitem{Kim2} W. Y. Kim and K. S. Kim, Acc. Chem. Res. {\bf 43}, 111 (2010).

\bibitem{JWYang} J. W. Yang, G. Lee, J. S. Kim, and K. S. Kim, J. Phys. Chem. Lett. {\bf 2}, 2577 (2011).

\bibitem{Wang} Z. Wang, C. Tang, R. Sachs, Y. Barlas and J. Shi, Phys. Rev. Lett.  {\bf 114}, 016603 (2015).

\bibitem{Leutenantsmeyer} J.C. Leutenantsmeyer, A.A. Kaverzin, M. Wojtaszek and B.J. van Wees, arXiv:1601.00995 (2016).

\bibitem{Evelt} M. Evelt, H. Ochoa, O. Dzyapko, V. E. Demidov, A. Yurgens, J. Sun, Y. Tserkovnyak, V. Bessonov, A. B. Rinkevich, and S. O. Demokritov, arXiv:1609.01613 (2016).

\bibitem{Singh} S. Singh et al, arXiv:1610.08017

\bibitem{Yang} H. X. Yang, A. Hallal, D. Terrade, X. Waintal, S. Roche, and M. Chshiev, Phys. Rev. Lett. {\bf 110}, 046603 (2013).

\bibitem{Qiao2} Z. Qiao, W. Ren, H. Chen, L. Bellaiche, Z. Zhang, A. H. MacDonald and Qian Niu, Phys. Rev. Lett. {\bf  112}, 116404 (2014).

\bibitem{Vobornik} I. Vobornik, U. Manju, J. Fujii, F. Borgatti, P. Torelli, D. Krizmancic, Y. S. Hor, R. J. Cava, and G. Panaccione, Nano Lett. {\bf 11}, 4079 (2011).

\bibitem{Zanolli} Z. Zanolli, Sci. Rep. {\bf 6}, 31346 (2016).

\bibitem{Zollner} K. Zollner, M. Gmitra, T. Frank, and J. Fabian, Phys. Rev. B 94, 155441 (2016).

\bibitem{Mendes} J. B. S. Mendes, O. Alves Santos, L. M. Meireles, R. G. Lacerda, L. H. Vilela-Le\~ao, F. L. A. Machado, R. L. Rodr\'iguez-Su\'arez, A. Azevedo, and S. M. Rezende, Phys. Rev. Lett.  {\bf 115}, 226601 (2015).

\bibitem{Wei} P. Wei, S. Lee, F. Lemaitre, L. Pinel, D. Cutaia, W. Cha, F. Katmis, Y. Zhu, D. Heiman, J. Hone, J. S. Moodera and C.-T. Chen, Nat Mater doi:10.1038/nmat4603 (2016).

\bibitem{Kresse1} G. Kresse and J. Hafner, Phys. Rev. B {\bf 47}, 558 (1993).
\bibitem{Kresse2} G. Kresse and J. Furthm\"uller, Phys. Rev. B {\bf 54}, 11169 (1996).
\bibitem{Kresse3} G. Kresse and J. Furthm\"uller, Comp. Mater. Sci. {\bf 6}, 15 (1996).
\bibitem{Blochl} P. E. Bl\"ochl, Phys. Rev. B {\bf 50}, 17953 (1994).

\bibitem{Perdew} J. P. Perdew, K. Burke, and M. Ernzerhof, Phys. Rev. Lett. {\bf 77}, 3865 (1996).
\bibitem{Kresse4} G. Kresse and D. Joubert, Phys. Rev. B {\bf 59}, 1758 (1999).

\bibitem{Soler}J. M. Soler, E. Artacho, J. D. Gale, A. Garc\`{i}a, J. Junquera, P. Ordej\'{o}n and Daniel S\'{a}nchez-Portal, J. Phys. Condens. Matter {\bf 14}, 2745 (2002).

\bibitem{Ceperley} D. M. Ceperley and B. J. Alder, Phys. Rev. Lett. {\bf 45},566 (1980).
\bibitem{Perdew2} J. P. Perdew and A.Zunger, Phys. Rev B {\bf 23}, 5075 (1981).

\bibitem{Anisimov1} I. Anisimov, J. Zaanen, and O. K. Andersen, Phys. Rev. B {\bf 44}, 943 (1991).

\bibitem{Anisimov2} I. Anisimov, I. V. Solovyev, M. A. Korotin, M. T. Czyzyk, and G. A. Sawatzky, Phys. Rev. B {\bf 48}, 16929 (1993).

\bibitem{Larson} P. Larson and W. R. L. Lambrecht, J. Phys.: Condens. Matter {\bf 18}, 11333 (2006).

\bibitem{Cho} S. B. Cho and Y. C. Chung, J. Mater. Chem. C, {\bf 1}, 1595 (2013).

\bibitem{Su} S. Su,  Y. Barlas and R. K. Lake, 	arXiv:1509.06427 (2015).

\end{thebibliography}
\end{document}